\documentclass[journal]{rmaa}
\usepackage{paralist}
\usepackage{amsmath}
\usepackage{psfrag,color}
\usepackage[usenames,dvipsnames]{xcolor}
\definecolor{afblue}{rgb}{0.36, 0.54, 0.66}
\definecolor{bronze}{rgb}{0.8, 0.5, 0.2}

\def\spose#1{\hbox to 0pt{#1\hss}}
\def\lta{\mathrel{\spose{\lower 3pt\hbox{$\mathchar"218$}}
     \raise 2.0pt\hbox{$\mathchar"13C$}}}
\def\gta{\mathrel{\spose{\lower 3pt\hbox{$\mathchar"218$}}
     \raise 2.0pt\hbox{$\mathchar"13E$}}}
\usepackage[normalem]{ulem}
\usepackage{url}
  
\title{Halo Occupation Distributions of Moderate X-ray AGNs \\ 
       through Major and Minor Mergers in a $\Lambda$-CDM Cosmology} 
\author{
L. Altamirano-D\'evora,
 T. Miyaji,
 H. Aceves,
 A. Castro,
 R. Ca\~nas,
 and F. Tamayo
\affil{Instituto de Astronom\'ia, Universidad Nacional
   Aut\'onoma de M\'exico, Ensenada, Baja California, M\'exico} 
   }

\shortauthor{Altamirano-D\'evora et al.}
\shorttitle{HOD of AGNs}
\fulladdresses{

\item L. Altamirano-Devora: Instituto de Astronom\'{\i}a, Universidad Nacional Aut\'onoma de M\'exico, Km 103 Carretera Tijuana-Ensenada, Codigo Postal 22860, Ensenada, B.~C, M\'exico (lili@astrosen.unam.mx).}

\listofauthors{L. Altamirano-Devora, T. Miyaji, H. Aceves, A. Castro, R. Canas \& F. Tamayo}

\indexauthor{Altamirano-D\'evora, L.}
\indexauthor{Miyaji T.}
\indexauthor{Aceves H.}
\indexauthor{Castro A.}
\indexauthor{Ca\~nas R.}
\indexauthor{Tamayo F.}

\abstract{Motivated by recent inferred form of the halo occupation distribution (HOD) of X-ray selected AGNs, in the COSMOS field by \citet{allevato12}, we investigate the HOD properties of moderate X-ray luminosity Active Galactic Nuclei (mXAGNs) using a simple model based on merging activity between dark matter halos (DMHs) in a $\Lambda$CDM cosmology. The HODs and number densities of the simulated mXAGNs at $z=0.5$, under the above scenarios to compare with \citet{allevato12} results. We find that the simulated HODs of major and minor mergers, and the observed for mXAGNs are consistent among them. Our main result is that minor mergers, contrary to what one might expect, can play an important role in activity mAGNs.}

\resumen{Motivados por la forma inferida de la Distribuci\'on de Ocupaci\'on de Halos (HOD) de N\'ucleos Activos de Galaxias (AGNs), seleccionados de rayos X en el campo de COSMOS por \citet{allevato12}, investigamos la HOD de N\'ucleos Activos de Galaxias de rayos X moderados (mXAGNs) usando un modelo basado en actividad por fusion entre halos de materia oscura (DMHs) en una cosmolog\'ia $\Lambda$CDM. La HOD y las densidades num\'ericas de los mXAGNs simulados en $z=0.5$, en los escenarios anteriores son calculados y comparados con los resultados de \citet{allevato12}. Encontramos un comportamiento similar entre las HODs simuladas de fusiones mayores y menores, y la observada para los mXAGNs. El resultado principal es que las fusiones menores, contrario a lo que se podria esperar, pueden jugar un papel importante en activar los mAGNs.}

\addkeyword{Galaxies: active}
\addkeyword{Methods: numerical, semi-analytical, N-body simulations}

\begin{document}
% Typeset article header
\maketitle
%%%%%%%%%%%%%%%%%%%%%%%%%%%%%%%%%%%%%%%%%%%%%%%%%%%%%%%%%%%%%%%%%%%%
%%%%%%%%%%%%%%%%%%%%%%%%%%%%%%%%%%%%%%%%%%%%%%%%%%%%%%%%%%%%%%%%%%%%

\section{Introduction}
\label{sec:intro} 

Studies of AGN host galaxies are fundamental to understand the physical mechanisms that trigger AGN activity and govern the fuelling rate of the central black hole \citep[e.g.,][]{Gilmour2009,Alex2012,BeckmannSrader2012}.

 The observed correlation between the mass of the central BH and the velocity dispersion ($\sigma$) of the bulge of the host galaxy suggest a strong connection between galaxy evolution and BH activity \citep[e.g.,][]{Geb2000, MeFa2001, tre2002, Ko2013}. This activity is related to the accretion of material to the central engine triggered by, for example: the merger of gas-rich galaxies \citep[e.g.,][]{SiRe1998, spr2005,Hop2008}, bar-driven inflows \citep[e.g.,][]{jog2006}, disk-instabilities \citep{Bou2011}, collisions with molecular clouds \citep{HoHe2006}, stellar winds from evolved stars \citep{CiOs2007}, and the transportation of gas to the centers by supernova explosions \citep{Ch2009} or a combination of these effects.

Mergers and strong interactions can induce substantial gravitational torques on the gas content of a galaxy, depriving it of its angular momentum, leading to inflows and the buildup of huge reservoirs of gas in its center \citep[e.g.,][]{He1989, BaHe1991, BaHe1996, MiHe1996, spr2005, Co2006, Co2008, DiM2007}. Major galaxy mergers are very efficient in moving gas to galaxy centers due to the generation of large torques. Observationally, visual inspections of host galaxies of AGNs \citep{Trei2012} find that the more luminous AGNs show recent merger features, while such features are not commonly seen in the less luminous AGNs, which appear to be driven by another process. Low luminosity AGNs could be triggered also in non-merger scenarios \citep[e.g.,][]{Milo2006, HoHe2006, HoHe2009}, and probably also by the interaction with very small satellites (total mass ratio of about 1:100) as recently suggested \citep{fox2014}.

The dominant process that triggers AGN activity could be a function of redshift and/or halo mass. The anti-hierarchical evolution of AGNs (or AGN down sizing), where the number density of low luminosity AGNs comes later in the universe than high luminosity ones \citep{ueda03,hasinger05,ueda14}, is probably not consistent with the theoretical predictions of a major merger AGN triggering scenario as suggested by \citet{wyithe_loeb03}. Some recent theoretical studies based on cosmological simulations suggest the necessity of a combination of merger and secular processes \citep[e.g.,][]{Draper2012} or hot-halo accretion and star-burst induced triggering \citep{Fani2012} to explain the evolution of the luminosity function of AGNs and even their clustering properties \citep{Fani2013}. These results suggest that one or more mechanisms other than major mergers are at least partially responsible for triggering the AGN activity.

Large scale AGN bias measurements show that mXAGNs are on average associated with more massive DMHs than more luminous QSOs \citep[e.g.,][]{miyaji07,krumpe10,allevato11}. While the typical masses of DMHs associated with QSOs [$M_\mathrm{DMH}\sim 10^{12-13} h^{-1}\mathrm M_{\sun}$; \citet{Po2004, Croom2005, Hop2007, Coil2007, da2008, Mou2009}] are consistent with a major merger triggering scenario \citep[e.g.,][]{shen09}, those associated with mXAGNs are typically more massive with $M_\mathrm{DMH}\sim 10^{13-14} h^{-1}\mathrm M_{\sun}$ (i.e., the mass scale of rich groups--poor clusters). While results from \citet{allevato11} suggest that secular processes could trigger mXAGNs. 

Cosmological simulations provide an important tool to understand the dark matter distribution in the universe, the co-evolution and growth of BHs with respect to their host galaxies \citep[e.g.,][]{Sij2007, DiM2008, Thac2006}. In $N$-body simulations the requirement of relating the dark matter to galaxy distributions has to be satisfied \citep[e.g.,][]{PuGa2014}. Studies of AGN clustering using cosmological simulations have been carried out using the halo model \citep[e.g.,][]{Thac2009, Deg2011} or with the BH continuity equation approach \citep[e.g.,][]{Lid2006, Bono2009, Sha2010}. The HOD method allows us to distinguish among AGN evolution models \citep[e.g.,][]{Chatte2012}. It has been used by several authors to interpret AGN and quasar clustering measurements from direct counts of AGNs within groups of galaxies \citep[e.g.,][]{Wake2008, shen2010, Miy2011, Stari2011, krumpe12, allevato12, Richa2012, KayOgu2012, Chatte2013,krumpe14}.

The environment of AGN, and in particular the mass of the typical DMHs in which they live, is a powerful diagnostic of the physics that drive the formation of super massive black holes (SMBHs) and their hosts galaxies \citep[e.g.,][]{MouGeo2012}. By modeling the mean AGN occupation at $z=0.5$, \citet{allevato12} found that the host halos of these AGNs have a DMH with mass $M_\mathrm{DMH}\geq10^{12.75} h^{-1}\mathrm{M_\odot}$, that it is a DMH mass for galaxy groups \citep{Eke2004}. This result agrees with studies from \citet{Geo2008} and \citet{Ar2009} that present evidence that AGNs at $z\approx1$ are frequently found in groups. On other hand, \citet{Miy2011} calculated the shape of the HOD of X-ray selected AGN that suggests that the AGN satellite fraction increases slowly with $M_\mathrm{DMH}$, in contrast with the satellite's HOD of low luminosity-limited samples of galaxies. For the latter galaxies, \citet{allevato12} found that the slope $\alpha$ of the HOD distribution of satellite AGNs had a value of $\alpha_{s}\leq 0.6$ in their analysis, suggesting a picture in which the average number of satellite AGNs per halo mass decreases with halo mass.

Considering that $\alpha\approx 1$ is inferred for galaxies in general \citep[e.g.,][]{coil2009, Ze2011}, the HOD studies of AGNs suggest that AGN fraction among galaxies decreases with increasing the DMH mass. The reason for this may be that the cross-section of merging between two galaxies decreases with increasing relative velocity and thus merging frequency is suppressed in a group/cluster environment with high velocity dispersion \citep{makino_hut97}. Also gas processes such as ram pressure stripping of cold gas in galaxies by hot intragroup/intracluster gas suppresses star formation activities that may feed AGN \citep{gunn_gott72}. 

In this work we investigate a scenario where satellite subhalos, from $N$-body cosmological simulations harbor mXAGNs within a group/cluster sized parent halo of $M_\mathrm{DMH}\geq10^{12.75} h^{-1}\mathrm{M_\odot}$ and have been triggered by either a major or minor merger. We compute the HOD of the simulated mXAGNs and compare them with the inferred HOD obtained by \citet{allevato12} and \citet{miyaji15} in order to see if we can reproduce such results by using this approach. In general we use a simple approach for coupling cosmological simulations $\Lambda$CDM with semi-analytical results to determine the HOD of our numerical AGNs.

The outline of the paper is as follows. In \S\ref{sec:model} we describe the method used in this work to determine AGN candidates in cosmological simulations to calculate their HOD. In \S\ref{sec:res} we show our results, discuss them in \S\ref{sec:dis}, and finally in \S\ref{sec:con} we indicate our main conclusions. Throughout this paper we adopt a matter density $\Omega_\mathrm{m}=0.266$, dark energy density $\Omega_\mathrm \Lambda=0.734$, $H_\mathrm{0}=72\mathrm{km\,s^{-1}\,Mpc^{-1}}$ and mass {\sc rms} fluctuation $\sigma_8=0.816$ consistent with the {\sc Wmap7} results of \citet{La2011}.

%%%%%%%%%%%%%%%%%%%%%%%%%%%%%%%%

\section{Model}
\label{sec:model}

In this section we describe the cosmological simulations used and semi-analytical procedure to relate the sub-halo satellites to observational properties of mXAGNs, then the criteria to determine when the merger occurs, and the actual computation of the HOD are described.

%%%%%%%%%%%%%%%%%%%%%%%%%%%%%%%%

\subsection{Numerical Simulations}
\label{sec:sim}

A set of five similar N-body cosmological simulations within the $\Lambda$CDM model, each differing from the others in the random seed used to generate the initial conditions were performed.

Each simulation box has a co-moving length of $L=100$$h^{-1}\,$Mpc with $N_ \mathrm p=512^{3}$ dark matter particles, each having a mass of $m_\mathrm{p}=6\times10^{8}$$h^{-1}\,\mathrm M_{\odot}$. Initial conditions were generated using 2nd-order Lagrangian Perturbation Theory \citep[e.g.,][]{croc2006} starting at a redshift of $z=50$. The initial linear power spectrum density was obtained from the cosmic microwave background code {\sc camb} \citep{Lew2000}. 

The $N$-body simulations were carried out using the publicly available parallel Tree$-$PM code {\sc Gadget}2 \citep{Springel2005}. The simulations were run with a softening length of $\varepsilon=20 \, h^{-1} \mathrm{kpc}$. Two cosmological simulations were resimulated with $\varepsilon=1 \, h^{-1} \mathrm{kpc}$ and no change in the HOD results was noted. The change in $\varepsilon$ can affect the properties of the inner profiles of halos, but that is out of the scope of the present paper.

%%%%%%%%%%%%%%%%%%%%%%%%%%%%%%%%

\subsection{Halo Finder Algorithm}
\label{sec:finder}

We identified DMHs and subhalos with the Amiga Halo Finder (AHF) code, which locates halos centers using an adaptive mesh refinement (AMR) approach. In brief, this code finds perspective halo centers, collects particles possibly bound to center, removes unbound particles and calculates halo properties \citep{Kno2009, Kne2011} \footnote{http://popia.ft.uam.es/AHF/Download.html}. Virial masses are defined using an overdensity of $200\rho_\mathrm{c}$, where $\rho_\mathrm{c}$ is the critical density of the universe. We used a minimum number of particles $N_\mathrm{p}=100$ to define a bound halo.

As mentioned in \S\ref{sec:intro}, we are interested in DMH with a virial mass of $M_\mathrm{DMH}\geq10^{12.75} h^{-1}\mathrm M_{\odot}\equiv M_{\rm th}$ at $z=0.5$ snapshot, in which mXAGN preferentially reside \citep{allevato12, Pad2009}. This is consistent with the HOD modeling of cross-correlation function between ROSAT all-sky Survey AGNs and luminosity red galaxies \citep{Miy2011}. In addition, we focus our attention to those that reside at a non-central location of the host halo. 

\begin{figure}[!t]
\centering
\includegraphics[width=6cm]{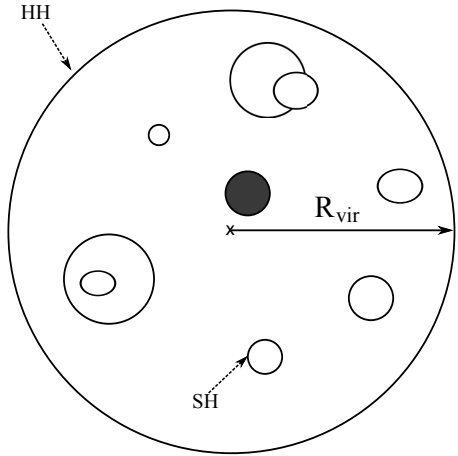}
\caption{Schematic diagram illustrates the Host$-$Halo (HH of mass $M_{\rm th}$) as circle with virial radius; $R_{\rm v}$. Several subhalos (SHs) are depicted inside $R_{\rm v}$. The central subhalo is identified as the closest to the center of the Host-Halo (dark filled circle) and the others are considered satellite subhalos (empty circles).}
 \label{fig:Halo}
\end{figure}

Halos with $M_\mathrm{DMH}>M_{\rm th}$ at redshift $z=0.5$ are called Host-Halos (HHs hereafter). We then identified subhalos that belong to these HHs, and selected the subhalos that are satellites (Subhalo-Host, SH), see Figure~\ref{fig:Halo}.

%%%%%%%%%%%%%%%%%%%%%%%%%%%%%%%%%%%%%%%%

\subsection{Connection subhalo to mXAGN}
\label{sec:cone}

Several approximate methods to assign AGN or quasar activity to halos are described in the literature. For example, \citet{croton09} uses the $M_\mathrm{BH}$-$\sigma$ relation by requiring to reproduce the observed luminosity function of quasars, using the abundance matching technique to ``turn-on'' halos in the {\sl Millennium} simulation \citep{smile2005}. \citet{CoW2013} invoked a model in quasars are treated as a light bulbs, through a empirical model for the demographics and M$_{BH}$-M$_{gal}$ to match the luminosity function of quasars. 

In this section we explain the different scaling relations used in our analysis, in order to relate the subhalos with observational properties of AGNs.

%%%%%%%%%%%%%%%%

	\subsubsection{Black Hole Mass}
	\label{sec:rela}

The central velocity dispersion $\sigma$ of each SH is related to the black hole mass $M_\mathrm{BH}$ \citep{Ko2013} by:

\begin{equation}
\label{eq:one}
\log\left(\frac{M_\mathrm{BH}}{M_0}\right)= -0{.}50 + 4.38 \log \left(\frac{\sigma}{\sigma_0}\right),
\end{equation}
where $M_0= 10^{9} h^{-1}\mathrm M_{\sun}$ and $\sigma_0= 200\, \mathrm{km}\,\mathrm s^{-1}$.

%%%%%%%%%%%%%%%%%%%%%%%%%%%%%%%%

	\subsubsection{Assigning Eddington ratio}
	\label{sec:lam}

Given a $M_\mathrm{BH}$, an Eddington ratio ($\lambda_\mathrm{Edd}$) can be defined as:

\begin{equation}
\label{eq:two}
\lambda_\mathrm{Edd} = \frac {L_\mathrm{bol}}{L_\mathrm{Edd}(M_\mathrm{BH})} \,
\end{equation}

where $L_\mathrm{bol}$ is the bolometric luminosity and $L_\mathrm{Edd}(M_\mathrm{BH})$ is the Eddington luminosity, which is proportional to $M_\mathrm{BH}$.

Combining equation \ref{eq:two} with the data provided in Table 2 of \citet{lusso12}, we can get the X-ray luminosity as:

\begin{equation}
\label{eq:three}
\log[L_\mathrm{bol}/L_\mathrm{band}] = a_{1} x+ a_{2} x^2 + a_{3} x^3 + b,
\end{equation}
where $L_\mathrm{band}$ correspond to the 0.5-2 keV band luminosity, x = $\log$ $L_\mathrm{bol}$$-$12, $a_\mathrm{1}$, $a_\mathrm{2}$, $a_\mathrm{3}$ and $b$ are bolometric correction coefficients.

We need to mimic a population of AGNs that represent mXAGNs within our simulations and construct the HOD to compare with the observed mXAGNs HOD by \citet{allevato12}. To do this we used a representative value of $\lambda_\mathrm{Edd}$=0.1, $a_\mathrm{1}$=0.248, $a_\mathrm{2}$=0.061, $a_\mathrm{3}$=-0.041 and $b$=1.431, restricting only subhalos with $L_\mathrm{x} \geq 10^{42.4} \, h^{-2}\,\mathrm{erg}\,\mathrm s^{-1}$.

Using equations~(\ref{eq:one}-\ref{eq:three}), we obtain the black hole mass threshold $M_\mathrm{\bullet}$, which we use to obtain the number density of subhalos with black holes that can be active or dormant, $n_\mathrm{(\geq M_\mathrm{\bullet})}$. We also consider a $z=0.5$ snapshot, which is the median redshift of the sample used by \citet{allevato12}. We will assign them to be active or dormant depending on whether they had suffered a galaxy merger within the AGN lifetime $\tau_\mathrm{AGN}$ in the past. 

%%%%%%%%%%%%%%%%%%%%%%%%%%%%%%%%

\subsection{Assigning AGN Lifetime }
\label{sec:life}

Instead of using the abundance matching technique, we select the duty cycle \citep[e.g.,][]{Cap2012} as an indicator that the sub-halo was turn on. In the following, we take a simple approach and assume that all AGNs observed at $z=0.5$ above the luminosity threshold, are shining at $\lambda_\mathrm{Edd}=0.1$ during their lifetime of $\tau_\mathrm{AGN}$.

Using the X-ray [2--10 keV] band luminosity function (XLF) of AGNs by \citet{miyaji15} and the model of the distribution function of the absorbing column density used in \citet{ueda14}, in combination with the results of \citet{allevato12}, we estimate the number density of AGNs including absorbed (within Compton-thin range, i.e., $N_{\rm H}< 10^{24}{\rm[cm^{-2}]}$) and the un-absorbed ones, above the intrinsic (i.e., before absorption) [0.5--2 keV] band luminosity of $L_\mathrm{x} \geq 10^{42.4}\, h^{-2}\,\mathrm{erg}\,\mathrm s^{-1}$. We obtain:

\begin{equation}
\label{eq:five}
n_\mathrm{AGN} \sim 4.2\times 10^{-5}h^3\mathrm{Mpc}^{-3}.
\end{equation}

The idea behind using the 2-10 keV luminosity function is that, the 0.5-2 keV sample used by \citet{allevato12} is highly selected against absorbed AGN. In order to estimate a more accurate AGN lifetime, we require a number density of both absorbed and un-absorbed AGNs. Here, for simplicity, we assume that the satellite HODs of absorbed and un-absorved mXAGNs have the same shape.

To constrain the number density of AGNs that we can observe at $z=0.5$ we have used the timescale $\tau_\mathrm{AGN}$, that indicates when the mXAGNs were activated. We estimate the AGN lifetime $\tau_\mathrm{AGN}$ as follows:
\begin{equation}
\label{eq:six}
\tau_\mathrm{AGN} \approx \frac{n_\mathrm{AGN}}{n_\mathrm{(\geq M_\mathrm{\bullet})}} \times \tau_\mathrm{age(z=0{.}5)},
\end{equation}
where $n_\mathrm{AGN}$ is the observed number density of Xray AGNs, $n_\mathrm{(\geq M_\mathrm{\bullet})}$ is the simulated number of SHs with BH mass threshold (active or dormant), and $\tau$$_\mathrm{age(z=0.5)}$ is the age of the universe at $z=0.5$.

%%%%%%%%%%%%%%%%%%%%%%%%%%%%%%%%

\subsection{Major and Minor Merger Criteria}
\label{sec:triggering}
 
We identify major and minor mergers at the redshift corresponding to the AGN lifetime $\tau_{\rm AGN}$ before $z=0.5$, so that the triggered AGNs by the mergers during this interval are still active at $z=0.5$ under these scenarios.

We define a mass ratio $\mu=M_\mathrm{2}/M_\mathrm{1}$ of the progenitors, where $M_\mathrm{2}$ $>$ $M_\mathrm{1}$. We consider those mergers with the mass ratios $0.25\leq \mu \leq 1.0$ as major and those with $0.1\leq \mu <0.25$ as minor mergers, respectively. In order to signal the merger event between two progenitors the following criteria between them have to be satisfied \citep[e.g.,][]{Fa1981}:

	\begin{enumerate}
          \item Their relative velocity $V_\mathrm{12} = \vert V_\mathrm{1} - V_\mathrm{2} \vert$ is less than average of velocity dispersion {\sc rms} of both halos $\langle V_\mathrm{rms} \rangle$; i.e., $V_\mathrm{12}\leq \langle V_\mathrm{rms} \rangle$.
	  \item Their relative physical separation $R_\mathrm{12}=\vert r_\mathrm{1}- r_\mathrm{2}\vert$ is less than the sum of the virial radius of both halos: $R_\mathrm{12} \leq R_\mathrm{v1} + R_\mathrm{v2}$.
	\end{enumerate}

To find these merger candidates, we used the {\tt MergerTree} tool which is included in the AHF software, we tag as progenitor the halo which contains the greatest fraction of SH particles \citep{Libeskind2010}. After this initial merging signaling, we verified by inspection of the snapshots that a merger event occurred.

%%%%%%%%%%%%%%%%%%%%%%%%%%%%%%%%

\subsection{Simulated HOD}
\label{hod}

The following formula is used in order to compute the HOD of mXAGNs:
	\begin{equation}
	\label{eq:seven}
	N(M_\mathrm{th}) = \frac{n_\mathrm{HH_\mathrm{agn}}} {n_\mathrm{HH}},
	\end{equation}
where $n_\mathrm{HH}$$_\mathrm{agn}$ is the number density of HHs that have a SH that has suffered a major or minor merger and has a $M_\mathrm{BH}$ $\geq$ $M_\mathrm{\bullet}$, and $n_\mathrm{HH}$ is the total number density of HH (defined in \S~\ref{sec:finder}) in simulations. The mass bin size used was $\Delta \log M_\mathrm{th}$ is 0.4.

%%%%%%%%%%%%%%%%%%%%%%%%%%%%%%%%%%%%%%%%%
%%%%%%%%%%%%%%%%%%%%%%%%%%%%%%%%%%%%%%%%%
\section{Results}
\label{sec:res}

Before presenting our results we make note of the following. The HOD of \citet{allevato12} was evaluated at $z=0$, and their HOD measurement made over a sample extending up to $z\sim 1$ that was corrected for the 0.5-2 keV XLF and its luminosity-dependent evolution, as indicated by \citet{ebrero09}. Their sample itself is more representative of $z\sim 0.5$ than $z\sim 0$.

\begin{figure}[!t]\centering
 \includegraphics[width=8cm]{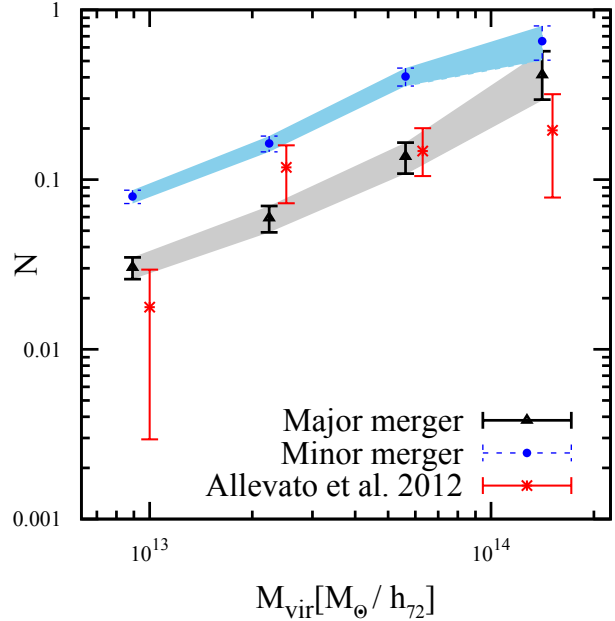}
 \caption{The HOD of the number of host that harbor a mXAGN triggered by either a major (blue dots) or minor merger (black triangles), and the inferred HOD of satellites mXAGNs by \citet{allevato12} (red asterisks). The latter corrected by an evolution factor and adding the un-absorbed AGN estimation of \citet{miyaji15}. Error bands are calculated as indicated in the text.}
\label{fig:fig3} 
\end{figure}

 To compare with our results, we back-correct their HOD to $z\sim 0.5$ in the following way. The Luminosity-Dependent Density Evolution (LDDE) model, describing the 0.5-2 keV XLF derived by \citet{ebrero09}, indicates that the number density grows as $\propto (1+z)^{3.38}$ up to $z\sim 0.8$ at all luminosities. Thus, we convert the $z=0$ HOD to $z=0.5$ HOD results by multiplying by $(1+0.5)^{3.38}=3.9$ at all DMH masses. Furthermore, since the \citet{allevato12} sample in the 0.5-2 keV band is highly selected against obscured AGNs, we further multiply the HOD by a factor of 2 to account for the obscured AGN contribution using the recent 2-10 keV XLF model by \citet{miyaji15}. In Figure \ref{fig:fig3}, we plot our results for major and minor mergers HOD and the corrected form of HOD to a redshift of $z=0.5$.

The slope of the HOD shows the same trend for both minor and major mergers for masses $\lta 5\times 10^{13} h^{-1} \rm{M}_\odot$. However, at higher masses the minor and major HODs display a somewhat different behavior; with the major merger HOD increasing and the minor one tending to be flat (see Figure~\ref{fig:fig3}). To compare with the slope ($\alpha_{s}$)found by \citet{allevato12}, we use the same model of occupation function, which is described by:
\begin{equation}
\label{eq:eight}
\langle N_\mathrm{sat}\rangle(M_\mathrm{h}) = f'_a \left(\frac{M_\mathrm{h}}{M_\mathrm{1}}\right)^{\alpha_{s}} \exp(- M_\mathrm{cut}/M_\mathrm{h});
\end{equation}
where $f'_{a}$ is a normalization, $M_\mathrm{1}$ is the halo mass at which the number of central AGN is equal to that of satellite AGNs ($\log M_{1}$=13.8M$_\mathrm{\sun}$) and $M_\mathrm{cut}$ is a cut-off mass scale ($\log M_\mathrm{cut}$=13.4 $M_\mathrm{\sun}$).

The fitted slope for the minor merger case was $\alpha_{s}$$=$ 0.10$\pm$0.09 and $\alpha_{s}= 0.20\pm0.18$ for the major merger, which can both be compared with the observed slope obtained by \citet{allevato12} $\alpha_{s}\leq 0.6$ (Table~\ref{tab:numden}). The slope of minor merger HOD is closer to that of the mXAGN HOD than that of the major merger case. However, both slopes are consistent with the observations.

\begin{table}[!t]\centering
	\scriptsize
 	\setlength{\tabnotewidth}{0.5\columnwidth}
  	\tablecols{3} 
  	\setlength{\tabcolsep}{2.8\tabcolsep}
  	\caption{Number Densities and HOD slopes} \label{tab:numden}
 	\begin{tabular}{lrr}
    	\toprule
   	Mechanism & \multicolumn{1}{c}{$n_\mathrm{agn}$} & \multicolumn{1}{c}{$\alpha$$_\mathrm{s}$} \\
    	\midrule
	Major & 2.28$\times$10$^{-5}$ & 0.20 $\pm$0.18 \\
	Minor & 5.96$\times$10$^{-5}$ & 0.10 $\pm$0.09\\
	Observed\footnotemark[1] & 4.2$\times$10$^{-5}$ & 0.22$^{+0.41}_{-0.29}$ \\
	\bottomrule
 \end{tabular}
\footnotemark[1]{\citet{allevato12}}
\end{table}

Errors in Figure~\ref{fig:fig3} were derived as follows. If $n_\mathrm{HH}$$_\mathrm{agn}$ is less than 15, we calculate 1$\sigma$ errors using equations (7) $\&$ (12) of \citet{geh1986}. If $n_\mathrm{HH}$$_\mathrm{agn}$ $\geq 15$, we calculate the 1$\sigma$ errors by $\sqrt{{n_\mathrm{HH}}_\mathrm{agn}}$.

%%%%%%%%%%%%%%%%%%%%%%%%%%%%%%%%%%%%%%%%%
%%%%%%%%%%%%%%%%%%%%%%%%%%%%%%%%%%%%%%%%

\section{Discussion}
\label{sec:dis}

Different numerical works have addressed the triggering of AGNs, in particular through mergers between galaxies since is a naturally expected contributing process \citep[e.g.,][]{san1988, Hop2006}. Major mergers are considered to activate QSO's, a situation that has been studied through hydrodynamical cosmological simulations, the measurements of their clustering and the properties of the AGNs \citep{Sij2007,DiM2008,marulli2009,ciotti2010,Deg2011,Chatte2012,van2012,krumpe15}.

However, \citet{Scha2011} found that most of the quasars in their sample have disk-like morphologies, suggesting that a secular evolution mechanism can drive the activity to this type of AGNs. Moreover, if only major mergers came to be important at high redshifts the AGNs should probably reside in more elliptical shaped galaxies \citep{Cis2011}. In contrast, \citet{Lee2012} and \citet{Cisternas2013} found that the gas bar-driven and the gas that could trigger nuclear activity do not correlate; a mechanism discussed in particular by \citet{wyse2004}.

Stochastic accretion models have been thought to be the triggering mechanism of low/moderate AGN \citep{HoHe2006,Hopkins2014,Koce2012}. Therefore, secular evolution \citep[e.g.,][]{Ehl2015} came to be an important scenario as well as minor mergers \citep{DeR1998, HeMi1995, Tani1999, Ken2003, HoHe2009, karo2014}, owing to the fact that injection of gas is recurrent and keeps the accretion continuous, that can explain the no-disk AGNs and can influence in some cases the growth of SMBH \citep[e.g.,][]{Kaviraj2014}. Considering that intermediate mergers are more common than major mergers \citep[e.g.,][]{Tapia2014}, these may also play a role. Recognizing the dominant triggering mechanism is not obvious in all types of AGNs, hence, we made a study in which mXAGNs are triggered by either major and minor mergers using a the HOD formalism and a simplified model. 

The accurate time of the merger is difficult to calculate, and therefore the time when the BH will be activated. If we use the dynamical friction time \citep[e.g.,][]{Hop2010} to estimate when the progenitors merge, we obtain an important overestimation in comparison to when we follow as closely as we can in time the evolution of the subhalos in our cosmological simulations. Furthermore, results by \citet{jiang2014} show that is not adequate to take a single timescale to infer when the merging takes place.

 In spite of the wide range of probable environments where AGN's live \citep{Villa2014, karo2014, leau2015}, in this work, we are assuming that the host/environment of the mXAGN have a group-like halo mass as indicated by some observations \citep[e.g.,][]{allevato12, sil2014}. We concentrated our attention in the mXAGNs that reside in non-central subhalos. The connection between galaxies and DMHs has been related in different ways: assigning the stellar mass of the galaxies to DMHs \citep{Deg2011, Beh2013}, introducing a gas fraction of galaxies into the DMHs \citep{Hop2010, Za2012} and using the luminosity function \citep{croton09}. Here we used a combination of semi-empirical and semi-analytic models to seed a black hole in a subhalo/satellite employing the relation $\sigma$-$M_\mathrm{BH}$.

The number densities of AGNs in group/cluster can help establish on firmer grounds whether there is any relation between environment density and AGN luminosity \citep{karo2014}, or even, the possible two-phase evolution in X-ray AGNs \citep{miyaji15}. It is shown in Figure~\ref{fig:fig3} that the selection of satellite subhalos have a similar distribution in the two merger scenarios tested, i.e., the shape of the HOD can be described to a good extent by both types of mergers. These processes also reproduce approximately the observed number densities of mXAGNs. The results shown in Figure~\ref{fig:fig3} indicate that minor mergers definitively play a role in establishing the HOD of these AGNs, and pin point to the necessity of further research along this line for igniting AGNs.

\begin{figure*}[ht]
\centering
 \includegraphics[width=2.1\columnwidth]{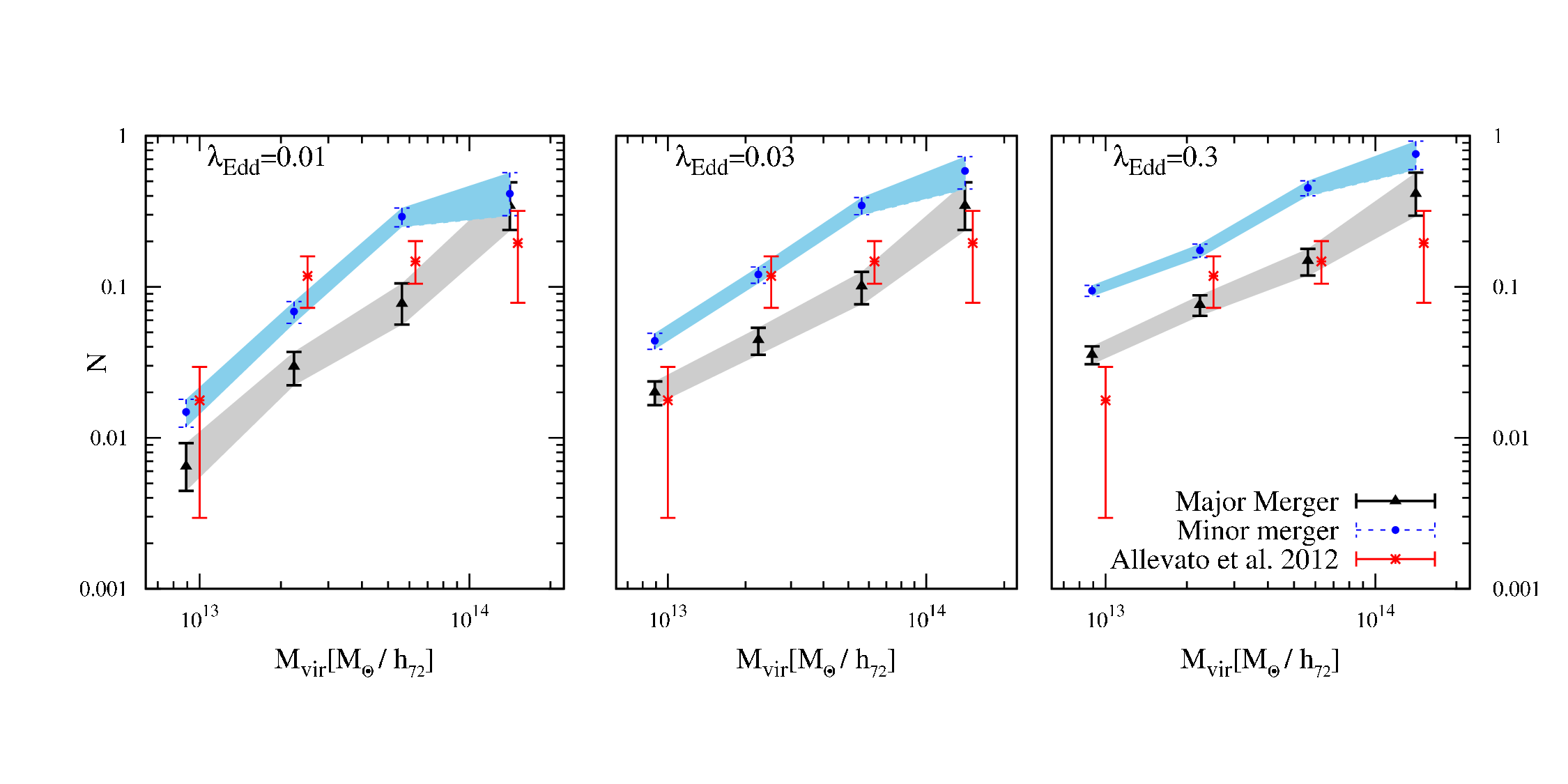}
 \caption{The HOD of different values of $\lambda$$_{Edd}$, the number of host that harbor a mXAGN triggered by either a major (blue dots) or minor merger (black triangles), and the inferred HOD of satellites mXAGNs by \citet{allevato12} (red asterisks).}
\label{fig:fig4} 
\end{figure*}

While we choose to use $\lambda_{\mathrm Edd}$=0.1 from the median value of the $\lambda_{\mathrm Edd}$ distribution from \citet{lusso12}, it is instructive to show how results change with $\lambda_{\mathrm Edd}$. Therefore we also computed simulated HODs for $\lambda_{\mathrm Edd}=0.01,0.03$ \& 0.3. The results are shown in Figure~\ref{fig:fig4} and $\alpha_{\mathrm s}$ values are shown in Table \ref{tab:lambda}. Figure~\ref{fig:fig4} and Table \ref{tab:lambda} show that the slopes are consistently flat for the minor merger case ($\alpha_{\mathrm s}\approx 0.1-0.2$) for all $\lambda_{\mathrm Edd}$ values, which are consistent with that of \citet{allevato12}. For the major mergers, the slope becomes steep ($\alpha_{\mathrm s}\approx 1$ for $\lambda_{\mathrm Edd}=0.01$), while it is flat for larger $\lambda_{\mathrm Edd}$ values. The global normalization of the minor merger HOD seems to match better with the observation for $\lambda_{\mathrm Edd}=0.01$ than higher $\lambda_{\mathrm Edd}$ values. However, we note that the normalization is directly proportional to the AGN lifetime $\tau_{\mathrm AGN}$ calculated from Eq. \ref{eq:six}, which is a rough approximation. Thus the agreement/disagreement of the HOD normalization between the models and the observation at a level of a factor of a few should not be used to prefer one model from another.

\begin{table}[!t]\centering
	\scriptsize
 	\setlength{\tabnotewidth}{0.5\columnwidth}
  	\tablecols{3} 
  	\setlength{\tabcolsep}{2.8\tabcolsep}
  	\caption{$\lambda$$_{\mathrm Edd}$ and $\alpha$$_{\mathrm s}$ } \label{tab:lambda}
 	\begin{tabular}{lrr}
    	\toprule
   	Lambda & \multicolumn{1}{c}{Major} & \multicolumn{1}{c}{Minor} \\
    	\midrule
	0.01 & 1.19 $\pm$0.16 & 0.18 $\pm$0.04 \\
	0.03 & 0.81 $\pm$0.19 & 0.11 $\pm$0.06 \\
	0.30 & 0.54 $\pm$0.20 & 0.15 $\pm$0.10 \\
	\bottomrule
 \end{tabular}
\end{table}

%%%%%%%%%%%%%%%%%%%%%%%%%%%%%%%%%%%%%%%%%%%%%%%%%%%%%%
%%%%%%%%%%%%%%%%%%%%%%%%%%%%%%%%%%%%%%%%%%%%%%%%%%%%%%
\section{Conclusions}
\label{sec:con}

In this work, we have used cosmological simulations and semi-analytical methods to assign activity to sub-halo satellites within halos with mass $M_\mathrm{th}\geq10^{12.75} h^{-1}\mathrm M_{\odot}$, and using the merger-driven scenario to trigger the BHs, to obtain an estimate of major and minor mergers contribution to the inferred shape of HOD of mXAGNs.

Testing different models can help to constrain the connection between the AGN and the host galaxy as well as the mechanism that triggers its BH. We have used an estimate of the duty cycle to relate the number density obtained from the simulate mXAGNs to the observed one. Our results bring forward the hypothesis that minor mergers, typically not considered as triggering events, can be an important factor in activating mXAGNs.

Our work shows that the HODs of mXAGNs under major merger and minor merger activation mode are very similar and both are consistent, in our approximate treatment, with the flat slope ($\alpha_s<1$) inferred form of HOD from observations. The minor merger model reproduces the slope of the satellite HOD of mXAGNs even at $\lambda$$_\mathrm{Edd}$=0.01, while that of the major merger model has a steeper slope at this low $\lambda$$_\mathrm{Edd}$.

On other hand, since our simulations do not take into account any baryonic processes, other mechanisms such as the ram-pressure stripping of cold gas is not excluded as a significant cause of the flat slope of the HOD. Our results, however, shows that non-baryonic processes such as the decrease of merging cross section in the high velocity encounters \citep{makino_hut97}, at least be able to produce a flat slope of satellite HOD.

With help of large X-ray surveys like eROSITA and larger samples of mXAGNs in group/cluster we will be able to constrain the properties of the host of these mXAGNs (e.g., mass), and more sophisticated models can lead us to get a better understanding of co-evolution of the AGNs and its distribution.

\section*{Acknowledgments}
%%%%%%%%%%%%%%%%%%%%%%%%%%%
 
This research was funded by UNAM-PAPIIT project IN108914, IN104113 and CONACyT Research Projects 179662. We thank Alexander Knebe for his help with the use of the AHF halo finder and thank Viola Allevato for the discussion of the values of the HOD, as well as to Vladimir Avila-Reese for helpful comments. 

%%%%%%%%%%%%%%%%%%%%%%%%%%%%%%%%%%%%%%%%%%%%%%%%%%%%%%%%%%%%%%%%%%%%
%%%%%%%%%%%%%%%%%%%%%%%%%%%%%%%%%%%%%%%%%%%%%%%%%%%%%%%%%%%%%%%%%%%%

\end{document}